\documentclass[reprint,superscriptaddress]{revtex4-1}

\pdfoutput=1
\usepackage[colorlinks,linkcolor=black,urlcolor=black,citecolor=blue]{hyperref}

\usepackage{graphicx}
\usepackage{amsmath}
\usepackage{multirow}
\usepackage{tabularx}

\begin{document}

\title{A Novel Analysis Method for Emission Tomography}

\author{Costas N. Papanicolas}
\email[Corresponding author: ]{papanicolas@cyi.ac.cy}
\affiliation{The Cyprus Institute, Konstantinou Kavafi 20,  2121 Nicosia, Cyprus}

\author{Loizos Koutsantonis}
\affiliation{The Cyprus Institute, Konstantinou Kavafi 20,  2121 Nicosia, Cyprus}

\author{Efstathios Stiliaris}
\affiliation{Physics Department, National and Kapodistrian University of Athens, Ilissia University Campus, 15771 Athens, Greece}
\affiliation{The Cyprus Institute, Konstantinou Kavafi 20,  2121 Nicosia, Cyprus}

\date{\today}

\begin{abstract}

We present a novel analysis method for image reconstruction in emission tomography.  The
method, named Reconstructed Image from Simulations Ensemble (RISE), utilizes statistical
physics concepts and Monte Carlo techniques to extract the parameters of a physical model
representing the imaged object from its planar projections.  Its capabilities are demonstrated
and evaluated by reconstructing tomographic images from sets of simulated SPECT projections.  The RISE results compare favourably to those derived from the well - known Maximum Likelihood Expectation Maximization (MLEM) method, the Algebraic Reconstruction Technique (ART) and the Filtered Back Projection (FBP).

\end{abstract}

%\pacs{}% insert suggested PACS numbers in braces on next line
\maketitle %\maketitle must follow title, authors, abstract and \pacs

\section{Introduction}
Emission tomography  is playing a dominant role in medical imaging with Positron Emission Tomography (PET) \cite{cherry:2001,vaquero:2015,van:2015, lecoq:2017} being the predominant modality followed by the Single Photon Emission Computerized Tomography (SPECT) \cite{ wernick:2004,madsen:2007,rahmim:2008, mariani:2010}. Other modalities such as Infrared Emission Tomography (IRET), relying on the same general principles are currently being investigated \cite{vavilov:2015}. 

In all emission tomography modalities the quality of the reconstructed images is limited by the background radiation emerging from the surrounding medium (e.g. tissue), absorption and re-scattering effects \cite{niu:2011,ritt:2011,lacroix:1994,Frey:1994,Koral:1988,Jas:1984}. In the case of PET and SPECT where the use of radiopharmaceuticals is required, and which are detrimental to the health of the patient, it is desired to achieve good image reconstruction while keeping the injected dose to a minimum. Minimizing the dose implies reconstruction from fewer images and/or with limited statistics.

In an effort to address the aforementioned considerations a novel method has been developed in which the 2D or 3D tomographic images are reconstructed from an ensemble of simulated solutions by statistically weighing the ones that satisfy a "goodness" criterion to the observed data. This method, in the interest of brevity, will be referred below as \textbf{\textit{"Reconstructed Image from Simulations Ensemble" (RISE)}}.

 RISE and the underlying  Athens Model Independent Analysis Scheme (AMIAS) method \cite{Papanicolas:2012} provide a highly complex and computationally intensive method for inverse problems which has proven successful in quantum inverse problems (scattering) in nuclear \cite{Markou:2018,Alexandrou:2012,Stiliaris:2007} and particle physics \cite{Alexandrou:2015}. It is based on statistical physics concepts and its implementation relies on Monte Carlo simulation techniques. It is a method of general applicability, well-suited for cases characterized by low statistics, noisy and attenuated data, as is often the case for SPECT and PET.

We present in subsequent sections the theoretical framework describing RISE and its computational implementation. We present its features in concrete examples in the case of SPECT modality. Software phantoms are employed to test the capabilities of the method and benchmark it versus the well established and well-understood techniques, the Maximum Likelihood Expectation Maximization (MLEM), the Algebraic Reconstruction Technique (ART) and the Filtered Back Projection (FBP).

\section{METHODOLOGY}

\subsection{The Tomographic Problem}

In emission tomography, the sectional image of an object from which radiation is emitted is reconstructed from projection  measurements  $\{Y_i|$ $i=1 \cdots NP\times NR\}$ obtained at different angles. $NP$ is the total number of projection angles and $NR$ is the number of bin (pixelated) measurements per projection angle. 

The prevailing practice in the field of medical imaging is to represent the  $N \times N$ pixelated tomographic image by the vector $\{F_j|j=1 \cdots N\times N\}$. The relation between the set of measured quantities $\{Y_i\}$ to the set of elements $\{F_j\}$ can be expressed as:
\begin{equation}
Y_i = \sum_{j=1}^{NP \times NR}{P_{ij}F_j}
\label{Eq:Proj}
\end{equation}
where $P_{ij}$ is a weighting matrix, also referred to as the projection matrix \cite{Epstein:2007} linking the image vector $F_j$ to the set of  projections $Y_i$.   $F_j$ denotes the intensity of the emitted radiation from the elemental area defined by the $j^{th}$ pixel. 

In the implementation of RISE presented in this paper, we do not examine issues of attenuation and scattering. The various approaches dealing with these issues can easily be implemented in RISE and indeed be further extended.

\subsection{Established Reconstruction Techniques}

The commonly used image reconstruction techniques in emission tomography are classified into two main categories: The first includes analytical methods such as the Filtered Back Projection (FBP) and its variants. These techniques, based on analytic inversion formulae \cite{Natterer:2001,Novikov:2002} of the Radon transform \cite{Radon:1917} and utilizing additional convolution kernels act on either the projection or image space, to provide images requiring minimal computational effort. The second category comprises the iterative methods that approximate the tomographic problem as a system of linear equations \cite{bruyant:2002,qi:2006}. These methods produce the reconstructed image following an iterative scheme for the minimization of a predefined cost function or maximization of the likelihood function.
 The most widely used iterative methods in emission tomography are the Algebraic based Reconstruction Technique (ART) \cite{Gordon:1970,Gilbert:1972} and the the Maximum Likelihood Expectation Maximization (MLEM) \cite{Shepp:1982,Lange:1984} including  its accelerated version, the Ordered Subsets Expectation Maximization (OSEM) method \cite{hudson:1994}. Compared to the  Filtered Back Projection (FBP), the iterative methods exhibit slower convergence but lead to more reliable reconstruction results especially in tomographic problems where a limited number of planar projections are available \cite{Wiez:2010,ma:2013,wolf:2013} and/or these projections are characterized by noise and attenuation.  
 
 In this study, we employed the traditional FBP, the Newton-Raphson version of ART \cite{Angeli:2009} and an open-source version of MLEM \cite{gursoy:2014} to produce reconstructions for comparison to the images produced by RISE.
 
\subsection{The Reconstructed Image from Simulations Ensemble (RISE) Technique}

The Reconstructed Image from Simulations Ensemble (RISE) technique approaches the tomographic problem in an entirely different way which is based on the Athens Model Independent Analysis Scheme (AMIAS) methodology. RISE obtains the image of the object (in 3D) or a 2D sectional image of it by constructing a model representing the radiation emission sources in the imaged volume. The model can be very general so as not to bias the outcome or specific and restrictive if it is desirable to incorporate prior knowledge. The model chosen to represent the imaged object is characterized by a number of parameters which are determined from the imaging data. In principle, the best model representation can be obtained from the data utilizing any “goodness of fit” method; we opted to employ the rather elaborate formulation of AMIAS which was formulated and successfully employed in addressing similar (inverse) problems.

\subsubsection{The Athens Model Independent Analysis Scheme (AMIAS)}

Methods based on AMIAS are applicable in problems in which the parameters of interest (e. g. the location and size of radiating "hotspots") are linked to the observable quantities (data) through a model.  They incorporate Monte Carlo techniques to simulate the generation of the observable quantities and as such, they require heavy computational resources. Prior knowledge can be incorporated endowing the scheme with Bayesian capabilities. In order to extract the desired parameter values from the data, the method employs an appropriate "goodness" criterion to quantify the comparison between the model simulated predictions to the observables; typically this is chosen to be the $\chi^2$ criterion:  

\begin{equation}
\chi^2 = \sum_{i}\frac{({\tilde{Y}}_i-Y_i)^2}{\epsilon_i^2}
\label{Eq:x2}
\end{equation}
where $\tilde{Y}_i$ is the predicted by the model quantity, $Y_i$ is the corresponding measured quantity and  $\epsilon_i$ is its associated uncertainty.

The method extracts the model parameters values and the corresponding uncertainties from a set of measured quantities (observables). The extracted parameters allow the reconstruction of the imaged object according to the model employed to represent the data.

\subsubsection{RISE: Implementing AMIAS in emission tomography}
\label{sec:RISE}
In AMIAS the tomographic imaging problem is approached in the following manner: the plane or volume to be imaged is described by a model whose parameters are derived via the AMIAS methodology. Unless prior knowledge is to be incorporated, It is necessary that the model has adequate flexibility to accommodate all shapes and intensity variations, that could be expected, and that it introduces no model bias. As AMIAS derives the parameters of the model with the maximum precision the data allow and with an evaluation of their uncertainties, it offers the added benefit of attributing a level of confidence to the parametrization of the derived result. 

\begin{figure}[h!]
\centering{
\includegraphics[width=1.0\columnwidth]{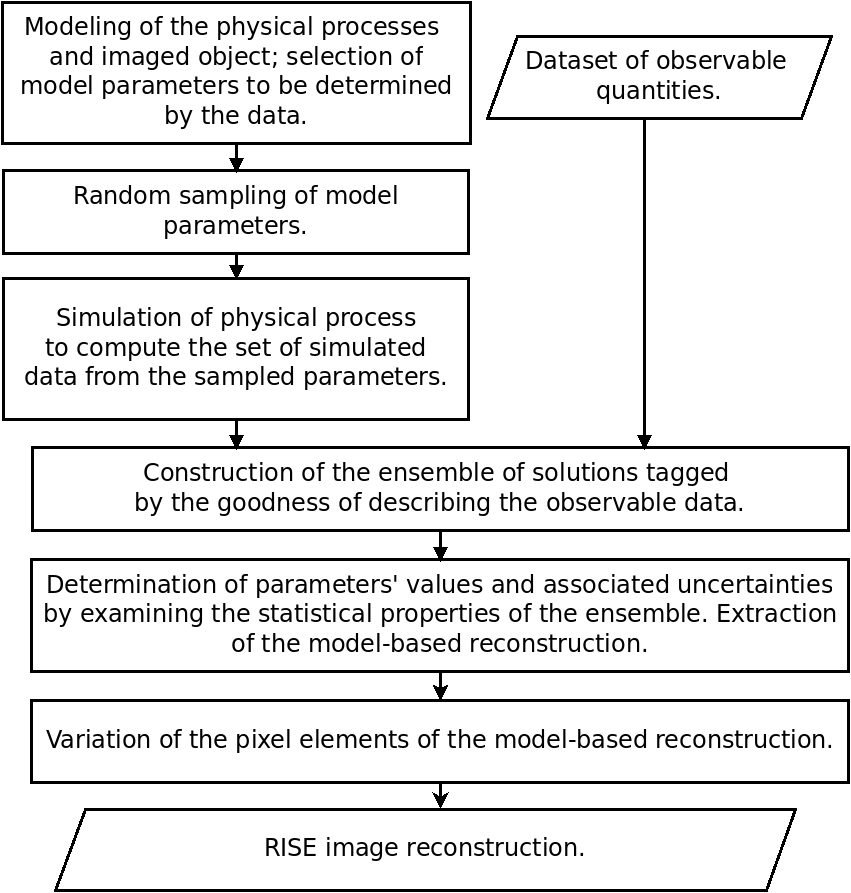}}
\caption{Flowchart of the algorithmic implementation of the AMIAS methodology\cite{Papanicolas:2012} in emission tomography. The parameters of a model describing the sources of radiation are randomly sampled in an iterative scheme. The simulated data are compared to the measured data to quantify the goodness of the simulated solution. From a sufficiently large ensemble of solutions, numerical results  are extracted for the parameters best describing the image by examining the statistical properties of the ensemble.}
\label{Fig:Amias}
\end{figure}

In RISE the  AMIAS methodology is implemented as follows (Figure \ref{Fig:Amias}):
\begin{enumerate}
\item A model is adopted in which the imaged distribution of the radiation sources is parameterized. In the variant presented here, the radiation distribution is described by a sum of elementary shapes representing the hotspots and a sum of smoothly and slowly varying terms representing the background.  Each elementary shape is specified by a set of parameters (e.g. position, size, intensity etc) and each background term by its amplitude. 
\item A random sampling procedure is used to choose a set of parameter values in a predefined range.  The set of randomly chosen parameters values uniquely and completely define the sources of radiation and as such, they provide a possible 'solution' to the problem; each set of model parameters corresponds to a given image reconstruction. In the choice of the model parametrization and of the ranges for the sampling of model parameters prior knowledge can easily be incorporated.
\item The tomographic image of the 'object' is constructed from the randomly chosen values of parameters. Projections of the 'object' are computed to simulate the measurement process incorporating the exact geometry of the imaging set-up and other processes such as the attenuation and scattering of photons in the intervening medium.
\item The simulated projections are compared to the measured projections by using the $\chi^2$ criterion (Eq: \ref{Eq:x2}). The $\chi^2$ value is assigned to each 'solution'  to quantify its 'goodness'  of representing reality, as captured in the experimental dataset.  
\item Steps $2-4$ are repeated (typically $10^4-10^6$ times) in order to construct a large ensemble of solutions. 
\item Numerical results for the parameters best describing the data (Probability Distribution Functions, mean values, and uncertainties) are extracted by examining the statistical properties of the ensemble. The derived parameters allow the reconstruction of the \textbf{\textit{model represantation}} of the imaged object.
\item The intensity of radiation from each individual pixel (voxel) of the model representation is allowed to vary to further improve the agreement of the reconstructed image with the data. This result in the  \textbf{\textit{RISE image reconstruction}}.
\end{enumerate}

\subsubsection{Modeling the Imaged Object}

\begin{figure*}[ht!]
\centering{
\includegraphics[width=\textwidth]{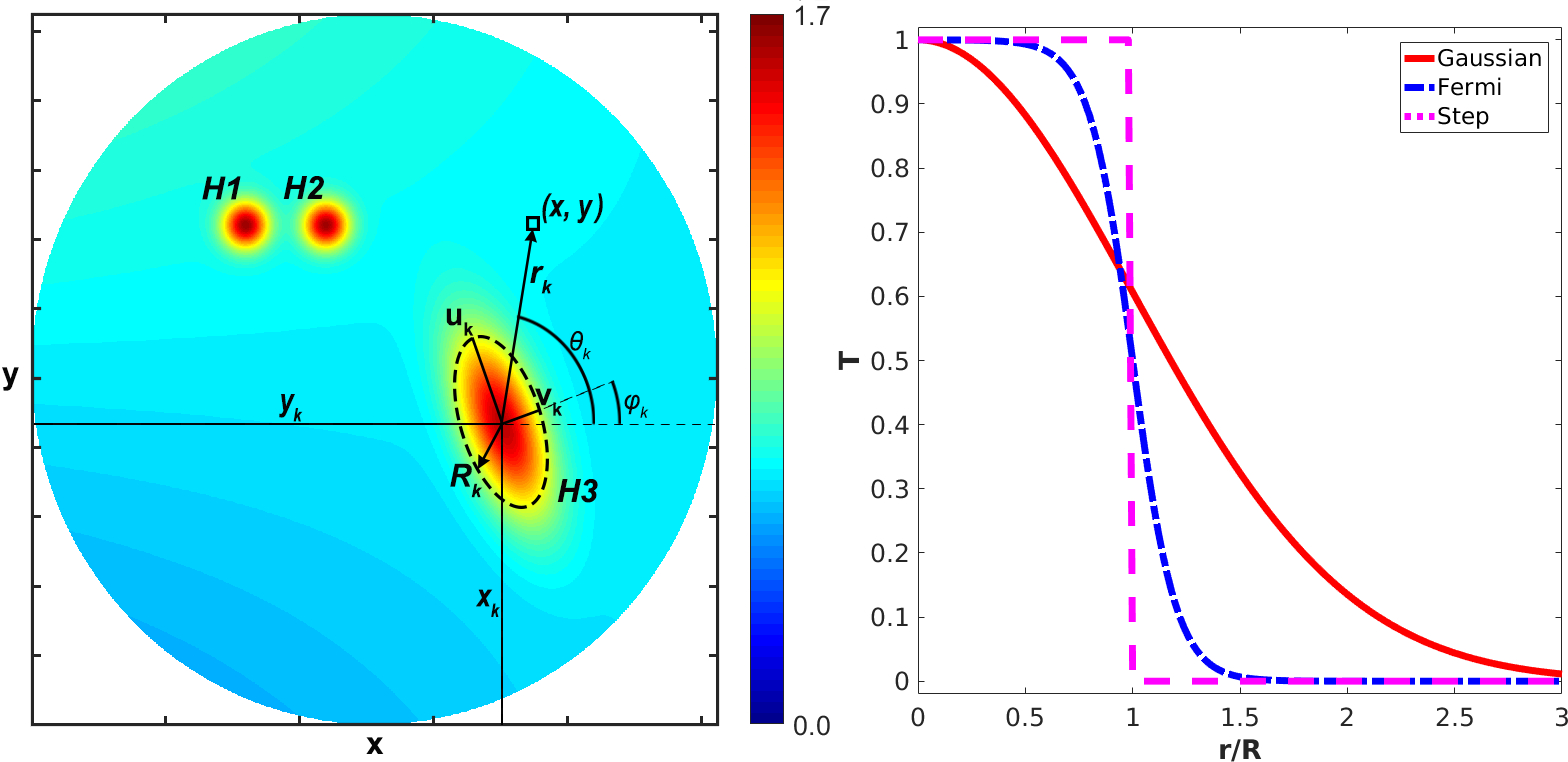}}
\caption{Left panel: Geometric representation of the 2D activity distribution $T(x,y)$ modeled according to Equation \ref{Eq:ModelTomo}; the geometric parameters in Equation \ref{Eq:Ellipse} describing the imaged object are depicted in the figure. The gaussian distribution (Equation \ref{Eq:ModelA}) was used to describe the activity intensity profile in each hotspot (indicated by H1-H3). Right panel: The radial profile of the three distribution functions used in modeling the profile of the radionuclide concentration in this paper.}
\label{Fig:Model}
\end{figure*}

In the case studies presented here, and in implementing the first step of the RISE algorithm (a model parametrization of the imaged object)  the 3D distribution of radiation sources is represented by a stack of  2D tomographic images, each one associated to a different offset in the third ("vertical") axis. The geometry of the activity distribution to be imaged in the 2D space is approximated by a set of fundamental shapes. In this work, these shapes are chosen to be ellipses which imply ellipsoidal shapes of radiation emitters in 3D. The parametric equation of an ellipse is given in the form of:
\begin{equation}
R(\theta) = \frac{u\cdot v}{\sqrt{(v \cdot cos(\theta-\phi))^2+ (u \cdot  sin(\theta - \phi))^2}}
\label{Eq:Ellipse}
\end{equation}
where $u$ and $v$ are the semi-major and semi-minor axes of the source ("hotspot") respectively and $\phi$ is the angle determining the orientation of the ellipse in the tomographic plane. The meaning of each parameter is best understood with the help of Figure \ref{Fig:Model}. The intensity of the radiation emitted by each point (pixel) in the volume specified defined by this elementary shape is described by an intensity profile function.

 The activity distribution in the imaged tomographic section is modeled by a sum of these elementary ellipses, each one (or a cluster of them) representing the presence of an emitting source ("hotspot"). Three different intensity profile functions providing the intensity distribution of the sources as a function of the geometrical factor $R$ are examined in this study.  All of them assume only radial dependence.

The first model ("Gaussian") assumes a Gaussian distribution:
\begin{equation}
T(x,y) = \sum_{k=1}^{N}{A_k \cdot exp{(-\frac{1}{2}\frac{r^2_{k}}{R^2_{k}})}}
\label{Eq:ModelA}
\end{equation}
where, $A_k$ is the  activity at the center of the $k^{th}$ "hotspot", $N$ is the number of "hotspots" and $r_{k}$ is the euclidean distance between the image element represented by its physical coordinates $(x,y)$ and the centroid $(x_k,y_k)$ of the $k^{th}$ "hotspot" (schematic illustration provided in Figure \ref{Fig:Model}). $T(x,y)$ is the activity value of the tomographic image element.

In the second  model ("Fermi"), the distribution of radioactivity is represented by the Fermi-like function:
\begin{equation}
T(x,y) = \sum_{k=1}^{N}{A_k}\cdot \Big(exp(\frac{r_{k}-R_{k}}{s_kR_k})+1\Big)^{-1}
\label{Eq:ModelB}
\end{equation}
where, in this model, an additional parameter $s_k$ (difussnes) is employed.

The third model ("step")  assumes a uniform activity distribution within the geometrical confines of the imaged hotspot and an abrupt change at its geometrical edges:
\begin{equation}
T(x,y) = \sum_{k=1}^{N}{A_k \cdot \Theta(R_{k}-r_{k}) }
\label{Eq:ModelC}
\end{equation}
where:
\[
 \Theta(R_{k}-r_{k}) =
  \begin{cases}
   1 , & \quad \text{for } R_{k} \geq r_{k}\\
   0 , & \quad \text{for }R_{k} < r_{k}\\
  \end{cases}\] 

The surrounding medium is also assumed to be an emitter of radiation having a simple slow varying spatial distribution:
\begin{equation}
B(x,y) = \sum^M_{i=0}C_iY_{i}(x,y)
\label{Eq:ModelBackg}
\end{equation}
where $Y_{i}(x,y)$ is a set of basis functions, $C_i$ are their amplitudes and $M$ is the number of radial terms chosen to describe the background distribution. In this study we describe the radiation emitted by the medium ("background") by using the set of Zernike polynomials $Z^j_{i}(x,y)$ \cite{len:2009,born:1959}. Thus:
\begin{equation}
B(x,y) = \sum^{M_z}_{i=0}\sum^i_{j=-i}C^j_iZ^j_{i}(x,y)
\label{Eq:ModelBackgZ}
\end{equation}
Thus the 2D imaged object is assumed to have an activity distribution $F(x,y)$ given by:
\begin{equation}
F(x,y) = T(x,y)+B(x,y)
\label{Eq:ModelTomo}
\end{equation}

\begin{figure*}[ht!]
\centering{
\includegraphics[width=\textwidth]{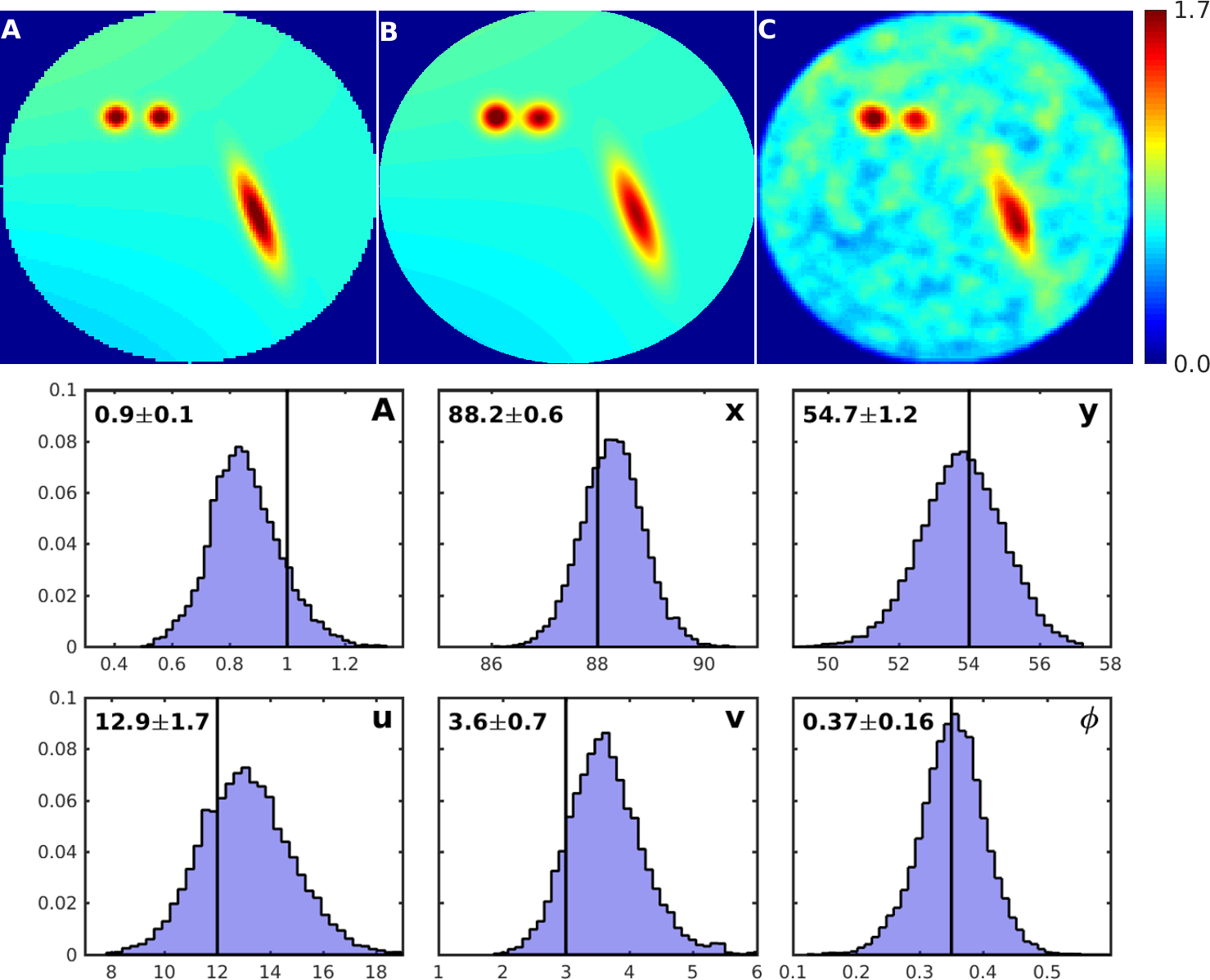}}
\caption{Upper panel: A) Software "Phantom A", the image of activity distribution simulated by a set of Gaussian functions (same as Figure \ref{Fig:Model}). B) The derived image providing the model representation of the phantom in which the parameters of interest are determined from their Probability Distribution Functions (PDF), tabulated in Table \ref{tab:parameters}. C) The  RISE reconstructed image of the phantom derived from the model representation using the last step of the RISE algorithm. Bottom panel: PDFs derived by RISE determining the parameters of the H3 "hotspot" in "Phantom A". Mean values and uncertainties calculated from the PDFs are shown at the top-left of the plots; the generator values used to simulate the phantom activity distribution are indicated by the solid lines. Values are shown in units of pixel-widths.}
\label{Fig:Phantom_Rec}
\end{figure*}   
  
\begin{table*}[ht!]
\renewcommand{\arraystretch}{1.7}
\caption{Parameters and associated uncertainties derived in RISE to produce the model representation of the phantom  shown in Figure \ref{Fig:Model}. The "true" values used to describe  the three hotspots H1, H2 and H3 and the background (B) are also shown. The empirical values and associated uncertainties for H3 are those derived from the PDFs shown in Figure \ref{Fig:Phantom_Rec}.
\label{tab:parameters} }
\begin{tabular*}{1.0\textwidth}{@{}l@{\extracolsep{\fill}}cccccc@{}}
\toprule
          & $A$             &  $x$  &  $y$ & $u$  & $v$ & $\phi$    \\ \hline 
          
H1        &  1.0             &   40.0           & 90.0            & 3.0 & 3.0 & 0.0          \\            
          & $1.1 \pm 0.3 $  & $40.7 \pm 0.9 $ & $89.8 \pm 0.7 $ & $3.5 \pm 0.9 $ & $3.5 \pm 0.8 $ & $0.1 \pm 0.8 $              \\ \hline               
          
H2        & 1.0             &  55.0           & 90.0            & 3.0 & 3.0 & 0.0          \\            
          & $0.9 \pm 0.3 $  & $55.4 \pm 1.2 $ & $89.4 \pm 0.9 $ & $4.0 \pm 1.7 $ & $3.7 \pm 1.5 $ & $0.0 \pm 0.8 $              \\ \hline

H3        & \ 1.0             &  88.0           & 54.0            &12.0 & 3.0 & 0.35          \\            
          & $0.9 \pm 0.1 $  & $88.2 \pm 0.6 $ & $54.7 \pm 1.2 $ & $12.9 \pm 1.7 $ & $3.6 \pm 0.7 $ & $0.37 \pm 0.16 $              \\ \hline \hline
  
          & $C^0_0$             &  $C^1_1$  &  $C^{-1}_1$ & $C^{-2}_2$  & $C^0_2$ & $C^2_2$     \\ \hline 
          
 B       & 0.6             &  0.08          & 0.0            & 0.04 &  0.0  & 0.0          \\            
       & $0.597\pm 0.004 $  & $0.07 \pm 0.01 $ & $0.005 \pm 0.007 $ & $0.033 \pm 0.007 $ & $0.003 \pm 0.008 $ & $0.001 \pm 0.009 $              \\             
          
\botrule   
   \end{tabular*}
\end{table*}

\subsubsection{Determining the model parameters}
\label{sec:DetParam}
 The RISE reconstruction process implements the algorithmic framework shown in Fig. \ref{Fig:Amias}. It allows the determination of an adequate number of background terms ($M$) and the number of "hotspots" ($N$)  which are needed to represent the imaged distribution (see Equation \ref{Eq:ModelTomo}). It results in the extraction of numerical values and associated uncertainties for the parameters  of the background and for each one of the $N$ modeled "hotspots".

RISE employs a fixed number of terms ($M$) in Equation \ref{Eq:ModelBackg} to represent the background distribution. High spatial frequency polynomials are excluded to avoid the spatial frequencies corresponding to the targeted "hotspots". Thus, the choice of a reasonably low number  $M$ prevents the correlation between the coefficients describing the higher frequencies of background $B(x,y)$  and the parameters describing the target $T(x,y)$.

 A number of algorithms were used to determine $N$. The easiest to explain is the one in which $N$ is determined by sequentially increasing the number of terms ($N=N+1$)  starting from $N=0$ till convergence is reached. For each increment of $N$, the Bayesian Information Criterion (BIC) \cite{schwarz:1978,neath:2012} is monitored and used to quantify the "goodness" of the model of $N$ terms:
 \begin{equation}
 BIC = \chi^2_{min}(N) + (n\cdot N+M) \cdot log{(NP \cdot NR)}
 \label{eq:bic}
 \end{equation}
 where $\chi^2_{min}(N)$ is the minimum $\chi^2$ value in the ensemble of solutions constructed for the model of $N$ terms, $n$ is the number of parameters describing each "hotspot", and $(NP \cdot NR)$ is the length of the sinogram. The optimum number of terms $N$ is selected as the one yielding the minimum BIC score which defines "convergence".

Having chosen the optimum number of terms ($N$) in Equation \ref{Eq:ModelTomo} the corresponding ensemble of solutions is used to determine the values of parameters for the model selected to describe the data.  This ensemble of solutions is then used to derive the Probability Distribution Function (PDF) of each one of the model parameters as specified by the AMIAS methodology \cite{Papanicolas:2012}. Mean values and higher-order moments are extracted from PDFs which provide the parameters' conventional optimum values.

 \subsubsection{Illustrative Case}
 
 To illustrate some key features of the method we apply the RISE methodology to reconstruct the $128 \times 128$ image of "Phantom A", shown in Figure \ref{Fig:Model} and panel A of Figure \ref{Fig:Phantom_Rec}. The phantom used comprises three "hotspots", two circulars in shape (H1 and H2) and a larger of ellipsoidal shape (H3); all three were given gaussian intensity profile. The background was described by the sum of the first six Zernike polynomials. Simulated data (sinograms) for 24 equidistant projections were generated using Equation \ref{Eq:Proj} which were subsequently randomized with Poisson noise.  A total of 24 parameters were used to describe the phantom.

   In applying the algorithm described in Section \ref{sec:RISE}, we modeled the object to be imaged by Equation \ref{Eq:ModelA} assuming a Gaussian intensity profile. Following the procedure of Section \ref{sec:DetParam} it was found that the phantom is best described by three ($N=3$) hotspots; the background was assumed to be adequately described by the sum of the first ten Zernike Polynomials ($M=10, M_z=3$). An ensemble of solutions was constructed comprising of 15000 simulated solutions. Employing the AMIAS methodology PDFs for each of the 28 free parameters of the model were derived in the $6^{th}$ step of the RISE algorithm which allows the model reconstruction of the phantom shown in panel B of the Figure \ref{Fig:Phantom_Rec}. 
   
   The derived PDFs for the position $(x,y)$, size $(u,v)$, orientation $(\phi)$ and intensity $(A)$ of hotspot H3 are shown in Figure \ref{Fig:Phantom_Rec}, along with the corresponding generator values. The values of the generator parameters and those extracted by RISE are shown in Table \ref{tab:parameters}, for all 24 input parameters of the model.  Excellent agreement is observed between the derived and generator parameter values. 
   
    The model representation is subsequently varied as prescribed in the  $7^{th}$ step of the RISE algorithm to yield the reconstructed tomographic image which is shown in panel C of the same figure.

\section{Simulation Studies and Evaluation of RISE} 

Software phantom simulations allowing direct similarity measures between the reconstructed images and the known true distribution were performed to evaluate and demonstrate the capabilities of the RISE method. Phantoms simulating the distribution of radiation sources were constructed on 2D grids and projection data were produced by using the forward projection model presented in Equation \ref{Eq:Proj}.

\subsection{Similarity Measures}
The assessment of the efficacy of a given reconstruction algorithm benefits greatly by the use of a quantitative criterion. For software phantoms the procedure is simple: the reconstructed distribution is compared to the known 'true' distribution by employing a merit function. Three well understood and widely used metrics, the Normalized Mean Square Error (NMSE), the Correlation Coefficient (CC) and the Peak Signal to Noise Ratio (PSNR) were used. 

The Normalized Mean Square Error (NMSE) provides a measure of the overall reconstruction error:
\begin{equation}
NMSE = \frac{\sum_{i=1}^{N^2}(F_i^{r}-F_i^{t})^2}{\sum^{N^2}_{i=1}{F^t_i}^2}
\label{eq:mse} 
\end{equation}
where $F^r_i$, $F^t_i$ are the pixel values, $N^2$ of them, of the reconstructed and phantom ("true") image respectively.
NMSE providing a normalized measure of the  absolute differences between the two images can be used in comparison studies exhibiting different activity ranges.

 The Correlation Coefficient (CC) allows similarity measures between two images without requiring the quantification of their absolute difference \cite{wolf:2013,Hill:2001}. 
CC is defined as:
\begin{equation}
CC \equiv \frac{\sum_{i=1}^{N^2}(F^r_i-\bar{F}^r)(F^t_i-\bar{F}^t)}{\sqrt{\sum_{i=1}^{N^2}(F^r_i-\bar{F}^r)^2\sum_{i=1}^{N^2}(F^t_i-\bar{F}^t)^2}}
\label{eq:corr}
\end{equation}
 where $\bar{F}^r$ and $\bar{F}^t$ are the mean values of the image elements $F^r_i$ and $F^t_i$ respectively.  
 Depending on the relation between the two images, reconstructed and phantom, the correlation coefficient attains values between $0.0$ and $1.0$. 

The Peak Signal-to-Noise Ratio (PSNR)  measured in decibels, is calculated by:
\begin{equation}
PSNR = 10\log _{10} \bigg(\frac{N^2 \cdot max(F^{r})}{\sum_{i=1}^{N^2}(F_i^{r}-F_i^{t})^2} \bigg)
\label{eq:cnr} 
\end{equation}
The higher the PSNR, the better the quality of the  reconstructed image.

\subsection{Software Phantoms}
\begin{figure}[ht!]
\centering{
\includegraphics[width=0.9\columnwidth]{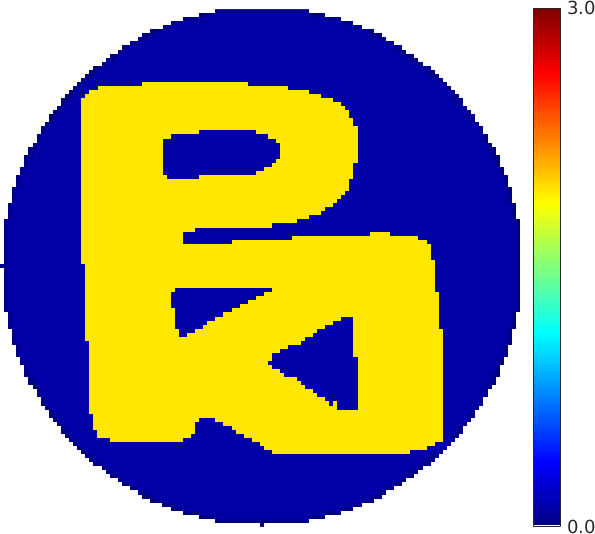}}
\caption{Software "Phantom B", a complex anagram with sharp edges,  used in simulation studies to validate the RISE  method.}
\label{Fig:Phantoms}
\end{figure}

 Two software phantoms were used to evaluate and demonstrate the capabilities of the RISE method:
\begin{itemize}
 \item "Phantom A" shown in Figure \ref{Fig:Model} is comprised of three hotspots of ellipsoidal shape exhibiting Gaussian intensity profile immersed in a non-uniform slow varying background activity. 
\item "Phantom B" shown in Figure \ref{Fig:Phantoms}, is a complex structure, an anagram, an overlay of the three Greek letters ($P$, $\Pi$, $K$) in a continuous uniform distribution with sharp (step) edges, placed in zero background. 
\end{itemize}
Although the methodology presented and the examples discussed are of general applicability to all modalities of emission tomography, the SPECT modality allows the closest realization of the simulation studies presented.

For each phantom, the image of the "true" activity distribution was sampled on a rectangular grid of 128 $\times$ 128 pixels size. Sets of vectorized projections (sinograms) were generated from the "true" images through Equation 1 by simulating 24 projections, evenly spaced in the full (360$^o$) angular range. The generated projections $\{Y^t_i,\quad i=1\dots NP\times NR\}$ obtained from the phantom images were further randomized with a Poisson probability distribution to provide the noisy sets of projections $\{\hat{Y}_i,\quad i=1\dots NP\times NR\}$:
\begin{equation}
\hat{Y}_i \sim Pois(Y^t_i)
\end{equation} 

\begin{figure*}[ht!]
\centering{
\includegraphics[width=\textwidth]{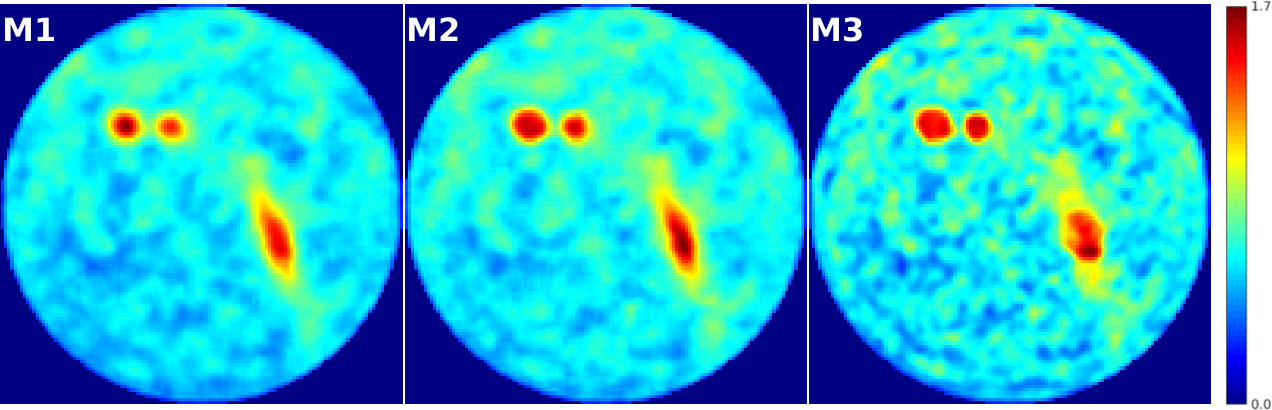}
}
\caption{Images of "Phantom A" reconstructed with the RISE method to evaluate the dependence of the reconstruction on the choice of the model that is used to represent the targeted hotspots in the tomographic plane. Three different models were used, the Gaussian (M1), the Fermi(M2) and the Step Function (M3).  }
\label{fig:Model_Recs}
\end{figure*}

The phantom pseudodata was used to examine three different aspects of the  RISE method:
\renewcommand{\labelenumi}{\Alph{enumi}}
\begin{enumerate}
\item{{\it Model Dependence:} To showcase the ability of RISE to reconstruct adequately the "true" activity distribution even in cases where the modeling of the elemental hotspots are only approximate. It is achieved by examining the dependence of the reconstruction on the choice of the intensity profile function for the targeted hotspots (Gaussian, Fermi and Step Function). Images of the "Phantom A" were reconstructed and compared.}
\item{{\it Shape Flexibility:} To demonstrate the ability of RISE  to reproduce any arbitrary shape especially in cases where the elemental hotspots employed do not resemble the geometrical characteristics of the imaged object. In this case study the image of "Phantom B" was reconstructed from its noisy projections by employing the series of ellipsoidally shaped sources.}
\item{{\it Minimization of Exposure} The ability of RISE to extract the "true" distribution from reduced statistics (in nuclear medical imaging corresponding to minimizing the exposure of the patient to radiation) is showcased. In a SPECT imaging simulation, "Phantom A" was used in three simulation cases (D1, D2, D3) to produce three sets of 24 projections. The data were generated by simulating 8048 (D1: full dose), 4028 (D2: half dose) and  2014 (D3: quarter dose) photon counts respectively and randomized with Poisson noise. Reconstructed images were obtained by using the Fermi model in the framework of RISE.  }
\end{enumerate}

 Reconstructions providing reference images for comparison were obtained in cases studies B and C  with MLEM, ART, and FBP.

\section{Reconstruction Results}

\subsection{Model Dependence}

Figure \ref{fig:Model_Recs} shows the reconstructed images of "Phantom A" as obtained in the first case study. 
The images were reconstructed  by using the three intensity profile functions: Gaussian, Fermi and step  defined by Equations \ref{Eq:ModelA}, \ref{Eq:ModelB} and \ref{Eq:ModelC} respectively.

\begin{table}[ht!]
\renewcommand{\arraystretch}{1.7}
\caption{CC, NMSE and PSNR values comparing the models used in RISE to reconstruct the image of "Phantom A". Values for both the model representation and the reconstructed (R) images are given.\label{tab:mod_dep} }
\begin{tabular*}{1.0\columnwidth}{@{}l@{\extracolsep{\fill}}ccc@{}}
\toprule
Model          &CC & NMSE & PSNR   \\ \hline
Gaussian       &0.94 & 0.029 & 24.86 \\ 
Gaussian (R)   &0.98 & 0.012 & 28.65\\ 
Fermi          &0.94 & 0.029 & 24.10 \\ 
Fermi (R)      &0.98 & 0.013 & 27.87\\ 
Step           &0.93 & 0.035 & 23.22 \\ 
Step (R)       &0.97 & 0.017 & 26.27\\      
\botrule   
   \end{tabular*}
\end{table}

A casual visual inspection of the reconstructed images shown in Figure \ref{fig:Model_Recs} reveals that the three images are very similar even in the case of the unrealistic step function distribution (M3).
This is borne out by the comparison of the corresponding CC, NMSE and PSNR values shown in Table \ref{tab:mod_dep}. The Gaussian model was expected to yield the best agreement as the same model was used to generate the "true" image of the phantom. Nevertheless, similar CC, NMSE and PSNR values to those of the Gaussian were obtained by using the Fermi model.  The lowest values were obtained for the step model. The step function was chosen to examine the ability of RISE to accommodate a manifestly deficient model which can not offer an acceptable representation of the phantom distribution. In such deficient model representation, the last step of the RISE algorithm proves to be important for the correction of the image. In the case of step function modeling, the CC value characterizing the reconstructed image was 1$\%$ lower than that obtained with the Gaussian model. It is demonstrated that the choice of a model for representing the intensity distribution is not critical for arriving at the correct outcome.

 \begin{figure*}[ht!]
\centering{
\includegraphics[width=\textwidth]{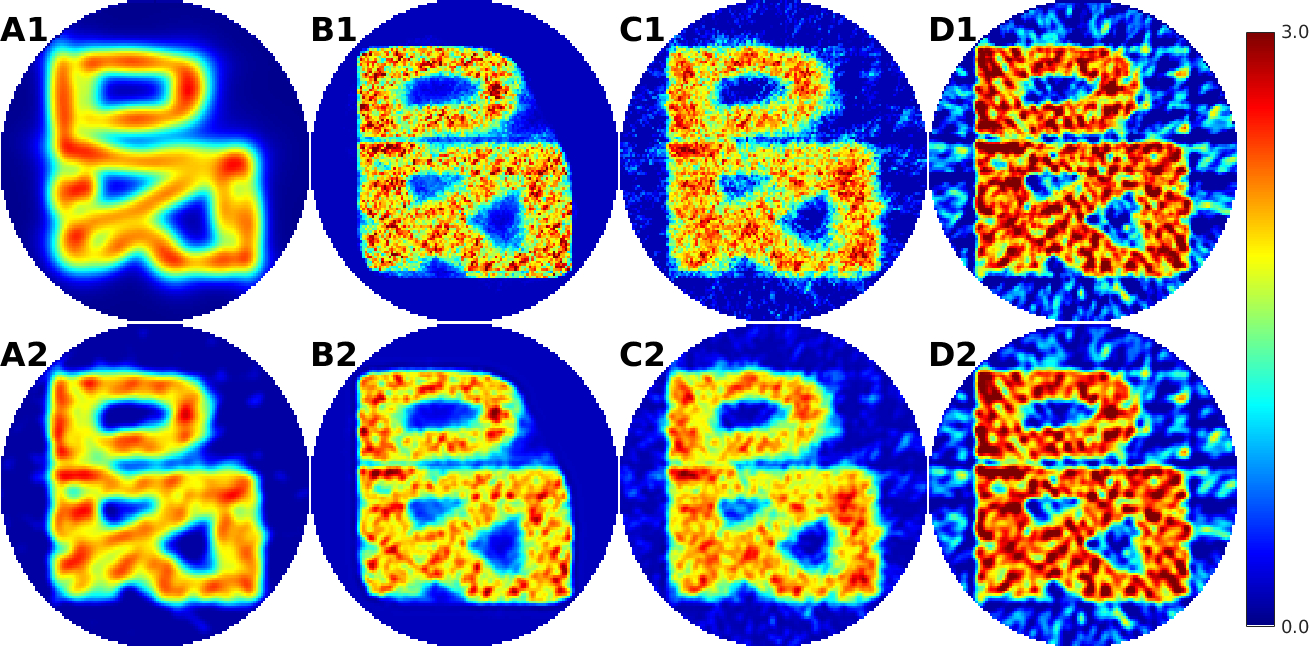}\\
}
\caption{Top row: The RISE, model representation (A1) and reconstructed (A2) images of "Phantom B" and the unfiltered images provided by MLEM (B1), ART (C1) and FBP (D1). Bottom row: The MLEM (B2), ART (C2) and FBP (D2) images post-filtered with a low-pass Butterworth filter are also shown. }
\label{fig:shape_app}
\end{figure*}
 
\subsection{Shape Approximation}
\label{sec:Shape}

RISE was applied to reconstruct the image of "Phantom B" by employing the summation of elementary ellipses having a Fermi activity profile as the approximating model. RISE model representation and reconstruction (A1 and A2 respectively)  images are shown in Figure \ref{fig:shape_app}.  Images reconstructed with MLEM (B1), ART (C1) and FBP (D1) are also shown for comparison. In the case of MLEM and ART, seven and two grand iterations were performed respectively.  MLEM, ART and FBP reconstructions were further post-filtered with a 3$^{rd}$ order Butterworth filter of 0.25 cycles per pixel cut-off frequency and are shown in panels B2, C2 and D2 respectively.

The reconstructed images of "Phantom B" were evaluated in terms of CC, NMSE, and PSNR; the results are shown in Table \ref{tab:shape_dep}. Three of the four methods (RISE, MLEM, ART) yield comparable CC, MSE and PSNR values. The result indicates that, in cases where the true distribution is totally unknown and has a complex structure, the choice of the  RISE method leads to similar reconstructions to those of MLEM and ART. FBP, affected by the angular undersampling (24 projections were used), produced images of lower CC and CNR values as compared to the other three methods.

\begin{table}[ht!]
\renewcommand{\arraystretch}{1.7}
\caption{CC, NMSE and PSNR values evaluating the reconstructions of "Phantom B" obtained from RISE, MLEM, ART and FBP. The evaluated image as shown in Figure \ref{fig:shape_app} is denoted in column "Panel".\label{tab:shape_dep} }
\begin{tabular*}{1.0\columnwidth}{@{}l@{\extracolsep{\fill}}cccc@{}}
\toprule
            & Panel  &CC & NMSE & PSNR   \\ \hline
RISE        & A1     &0.93 &0.087 &17.31 \\ 
RISE (R)    & A2     &0.94 &0.070 &18.63\\ 
MLEM         & B1    &0.92 &0.093 &21.08 \\ 
MLEM (PF)    & B2    &0.94 &0.072 &20.56\\ 
ART          & C1    &0.92 &0.105 &18.90 \\ 
ART (PF)     & C2    &0.93 &0.085  &18.64\\    
FBP          & D1    &0.88 &0.254 &17.78 \\ 
FBP (PF)      & D2   &0.89 &0.234 &17.74\\   
\botrule   
   \end{tabular*}
\end{table}

\subsection{Minimization of Exposure}

\begin{figure*}[ht!]
\centering{
\includegraphics[width=\textwidth]{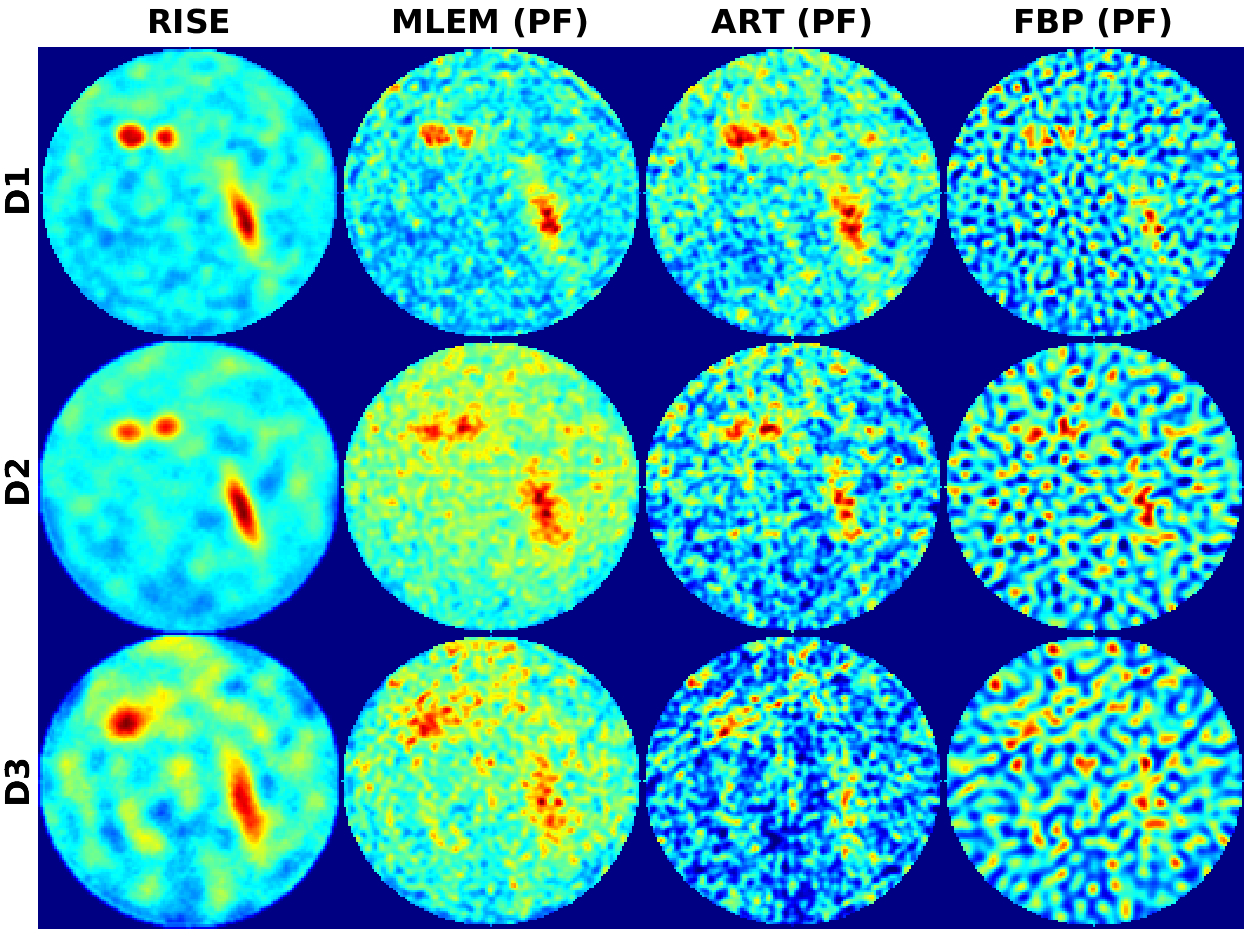}
}
\caption{Reconstructed images of "Phantom A" obtained from the simulated data in the three cases characterized by diminishing doses ($D_1=2D_2=4D_3$). The presented MLEM, ART and FBP image were post-filtered with a low-pass filter. The presented images were scaled to their
activity ranges in order to be visualized with the maximum possible contrast.}
\label{fig:Background}
\end{figure*}

Figure \ref{fig:Background} shows the reconstructed images of "Phantom A" as obtained with RISE, MLEM, ART, and FBP for the three simulation cases varying the number of simulated photons' counts (D1:8048 counts, D2:4024 counts, D3:2012 counts). RISE reconstruction was performed by employing the Fermi model.
MLEM, ART and FBP images were reconstructed and post-filtered as described in Section \ref{sec:Shape}.

\begin{table*}[ht!]
\renewcommand{\arraystretch}{1.7}
\caption{CC, NMSE and PSNR values calculated from the reconstructed images in the three simulation cases varying the number of simulated photons (images are shown in Figure \ref{fig:Background}). \label{tab:Background} }
\begin{tabular*}{1.0\textwidth}{@{}l@{\extracolsep{\fill}}|ccc|ccc|ccc@{}}
\toprule
        & \multicolumn{3}{c|}{D1 (Full Dose)}       &  \multicolumn{3}{c|}{D2 (Half Dose)}    &  \multicolumn{3}{c}{D3 (Quarter Dose)}      \\ 
        & CC & NMSE & PSNR & CC & NMSE & PSNR & CC & NMSE & PSNR     \\ \hline
        
RISE        &0.98 & 0.013 & 27.87 &0.97 & 0.017 & 26.17 & 0.92 & 0.041 & 21.28 \\      
             
MLEM (PF)   & 0.94 & 0.035 & 23.82 & 0.90 &0.050 & 19.61 & 0.86 & 0.068 & 18.77 \\    
            
ART  (PF)   & 0.90 & 0.055 & 21.15 & 0.80 & 0.127 & 18.34 & 0.71 & 0.203 & 17.80 \\
    
FBP  (PF)   & 0.80 & 0.402 & 16.87 & 0.75 & 0.433 & 15.77 & 0.73 & 0.497 &15.47\\
\botrule   
   \end{tabular*}
\end{table*}

A visual examination and comparison of the results reveal that the RISE image exhibits higher contrast and increased detectability of hotspots compared to those of MLEM. Compared to the images obtained with  ART and FBP, RISE images present less amount of noisy artifacts.

CC, PSNR and NMSE values evaluating the reconstructed images are shown in Table \ref{tab:Background}. In all simulation cases, RISE image scored the highest CC values followed by the post-filtered MLEM images. Furthermore, RISE led to an improvement in PSNR  as compared to the PSNR values of the three other methods.

\section{Specific applications}

Early results from on-going work applying the theoretical framework (RISE) presented in this paper have been recently reported for the evaluation and demonstration of the method in purpose-specific applications including experimentation with hardware phantoms. These studies include SPECT Dopamine Transporter Imaging (DAT) for Parkinson's disease \cite{koutsantonis:2017}, imaging of lungs and kidneys in mice \cite{koutsantonis:2017mouse} and extensive experimentation with $^{99m}$Tc phantoms utilizing the high-resolution SPECT camera of the University of Athens \cite{Spanoudaki:2004}. Moreover, the employment of the method in other modalities of emission tomography such as the Infrared Emission Tomography (IRET) is currently under evaluation. Results from an initial investigation of the RISE method in IRET, with attenuated and diffussed thermal data, can be found in a preliminary study\cite{koutsantonis:2016}.

\section{Summary Conclusions}
The theoretical foundation of a novel method, the Reconstructed Image from Simulations
Ensemble (RISE), for emission tomography, has been presented along with its algorithmic implementation. In the work presented here, RISE was demonstrated and evaluated with software phantoms in the SPECT modality resulting superior reconstruction results compared to the well established MLEM, ART, and FBP methods. 

The method is computationally demanding but robust and particularly well suited to noisy and low statistics data. The new method allows quality reconstructed data from fewer planar images and with lower statistics, thus allowing the use of a lower dose of a radiopharmaceutical in SPECT and presumably PET imaging, minimizing the exposure to the patient.

The RISE method, as presented here, provides a reconstruction framework of 2D tomographic images. The extension of the method to a 3D model providing the direct reconstruction of the volumetric image is straightforward and it is expected to provide further improvement to the 2D images; early results from the SPECT tomographic imaging of hardware phantoms appear to validate this expectation.

\begin{acknowledgements}
This  work was supported by the Graduate School of The Cyprus Institute and the Cy-Tera Project "NEA IPODOMI/STRATI/0308/31", which is co-funded by the European Regional Development Fund and the Republic of Cyprus through the Research Promotion Foundation. The authors would like to thank Theodoros Christoudias and Charalambos Chrysostomou for proofreading this manuscript.
\end{acknowledgements}

\section*{Conflicts of Interest}
The authors have no conflicts of interest, financial or other, to declare.

%\bibliography{bibliog.bib}

\begin{thebibliography}{45}%
\makeatletter
\providecommand \@ifxundefined [1]{%
 \@ifx{#1\undefined}
}%
\providecommand \@ifnum [1]{%
 \ifnum #1\expandafter \@firstoftwo
 \else \expandafter \@secondoftwo
 \fi
}%
\providecommand \@ifx [1]{%
 \ifx #1\expandafter \@firstoftwo
 \else \expandafter \@secondoftwo
 \fi
}%
\providecommand \natexlab [1]{#1}%
\providecommand \enquote  [1]{``#1''}%
\providecommand \bibnamefont  [1]{#1}%
\providecommand \bibfnamefont [1]{#1}%
\providecommand \citenamefont [1]{#1}%
\providecommand \href@noop [0]{\@secondoftwo}%
\providecommand \href [0]{\begingroup \@sanitize@url \@href}%
\providecommand \@href[1]{\@@startlink{#1}\@@href}%
\providecommand \@@href[1]{\endgroup#1\@@endlink}%
\providecommand \@sanitize@url [0]{\catcode `\\12\catcode `\$12\catcode
  `\&12\catcode `\#12\catcode `\^12\catcode `\_12\catcode `\%12\relax}%
\providecommand \@@startlink[1]{}%
\providecommand \@@endlink[0]{}%
\providecommand \url  [0]{\begingroup\@sanitize@url \@url }%
\providecommand \@url [1]{\endgroup\@href {#1}{\urlprefix }}%
\providecommand \urlprefix  [0]{URL }%
\providecommand \Eprint [0]{\href }%
\providecommand \doibase [0]{http://dx.doi.org/}%
\providecommand \selectlanguage [0]{\@gobble}%
\providecommand \bibinfo  [0]{\@secondoftwo}%
\providecommand \bibfield  [0]{\@secondoftwo}%
\providecommand \translation [1]{[#1]}%
\providecommand \BibitemOpen [0]{}%
\providecommand \bibitemStop [0]{}%
\providecommand \bibitemNoStop [0]{.\EOS\space}%
\providecommand \EOS [0]{\spacefactor3000\relax}%
\providecommand \BibitemShut  [1]{\csname bibitem#1\endcsname}%
\let\auto@bib@innerbib\@empty
%</preamble>
\bibitem [{\citenamefont {Cherry}(2001)}]{cherry:2001}%
  \BibitemOpen
  \bibfield  {author} {\bibinfo {author} {\bibfnamefont {S.~R.}\ \bibnamefont
  {Cherry}},\ }\bibfield  {title} {\enquote {\bibinfo {title} {Fundamentals of
  positron emission tomography and applications in preclinical drug
  development},}\ }\href@noop {} {\bibfield  {journal} {\bibinfo  {journal} {J
  Clin Pharmacol}\ }\textbf {\bibinfo {volume} {41}},\ \bibinfo {pages}
  {482--491} (\bibinfo {year} {2001})}\BibitemShut {NoStop}%
\bibitem [{\citenamefont {Vaquero}\ and\ \citenamefont
  {Kinahan}(2015)}]{vaquero:2015}%
  \BibitemOpen
  \bibfield  {author} {\bibinfo {author} {\bibfnamefont {J.~J.}\ \bibnamefont
  {Vaquero}}\ and\ \bibinfo {author} {\bibfnamefont {P.}~\bibnamefont
  {Kinahan}},\ }\bibfield  {title} {\enquote {\bibinfo {title} {Positron
  emission tomography: current challenges and opportunities for technological
  advances in clinical and preclinical imaging systems},}\ }\href@noop {}
  {\bibfield  {journal} {\bibinfo  {journal} {Annu Rev Biomed Eng}\ }\textbf
  {\bibinfo {volume} {17}},\ \bibinfo {pages} {385--414} (\bibinfo {year}
  {2015})}\BibitemShut {NoStop}%
\bibitem [{\citenamefont {Vandenberghe}\ and\ \citenamefont
  {Marsden}(2015)}]{van:2015}%
  \BibitemOpen
  \bibfield  {author} {\bibinfo {author} {\bibfnamefont {S.}~\bibnamefont
  {Vandenberghe}}\ and\ \bibinfo {author} {\bibfnamefont {P.~K.}\ \bibnamefont
  {Marsden}},\ }\bibfield  {title} {\enquote {\bibinfo {title} {{PET-MRI}: a
  review of challenges and solutions in the development of integrated
  multimodality imaging},}\ }\href@noop {} {\bibfield  {journal} {\bibinfo
  {journal} {Phys Med Biol}\ }\textbf {\bibinfo {volume} {60}},\ \bibinfo
  {pages} {R115} (\bibinfo {year} {2015})}\BibitemShut {NoStop}%
\bibitem [{\citenamefont {Lecoq}(2017)}]{lecoq:2017}%
  \BibitemOpen
  \bibfield  {author} {\bibinfo {author} {\bibfnamefont {P.}~\bibnamefont
  {Lecoq}},\ }\bibfield  {title} {\enquote {\bibinfo {title} {Pushing the
  limits in time-of-flight {PET} imaging},}\ }\href@noop {} {\bibfield
  {journal} {\bibinfo  {journal} {IEEE Trans Radiat Plasma Med Sci}\ }\textbf
  {\bibinfo {volume} {1}},\ \bibinfo {pages} {473--485} (\bibinfo {year}
  {2017})}\BibitemShut {NoStop}%
\bibitem [{\citenamefont {Wernick}\ and\ \citenamefont
  {Aarsvold}(2004)}]{wernick:2004}%
  \BibitemOpen
  \bibfield  {author} {\bibinfo {author} {\bibfnamefont {M.~N.}\ \bibnamefont
  {Wernick}}\ and\ \bibinfo {author} {\bibfnamefont {J.~N.}\ \bibnamefont
  {Aarsvold}},\ }\href@noop {} {\emph {\bibinfo {title} {Emission tomography:
  the fundamentals of {PET} and {SPECT}}}}\ (\bibinfo  {publisher} {Elsevier},\
  \bibinfo {year} {2004})\BibitemShut {NoStop}%
\bibitem [{\citenamefont {Madsen}(2007)}]{madsen:2007}%
  \BibitemOpen
  \bibfield  {author} {\bibinfo {author} {\bibfnamefont {M.~T.}\ \bibnamefont
  {Madsen}},\ }\bibfield  {title} {\enquote {\bibinfo {title} {Recent advances
  in {SPECT} imaging},}\ }\href@noop {} {\bibfield  {journal} {\bibinfo
  {journal} {J Nucl Med}\ }\textbf {\bibinfo {volume} {48}},\ \bibinfo {pages}
  {661--673} (\bibinfo {year} {2007})}\BibitemShut {NoStop}%
\bibitem [{\citenamefont {Rahmim}\ and\ \citenamefont
  {Zaidi}(2008)}]{rahmim:2008}%
  \BibitemOpen
  \bibfield  {author} {\bibinfo {author} {\bibfnamefont {A.}~\bibnamefont
  {Rahmim}}\ and\ \bibinfo {author} {\bibfnamefont {H.}~\bibnamefont {Zaidi}},\
  }\bibfield  {title} {\enquote {\bibinfo {title} {{PET} versus {SPECT}:
  strengths, limitations and challenges},}\ }\href@noop {} {\bibfield
  {journal} {\bibinfo  {journal} {Nucl Med Commun}\ }\textbf {\bibinfo {volume}
  {29}},\ \bibinfo {pages} {193--207} (\bibinfo {year} {2008})}\BibitemShut
  {NoStop}%
\bibitem [{\citenamefont {Mariani}\ \emph {et~al.}(2010)\citenamefont
  {Mariani}, \citenamefont {Bruselli}, \citenamefont {Kuwert}, \citenamefont
  {Kim}, \citenamefont {Flotats}, \citenamefont {Israel}, \citenamefont
  {Dondi},\ and\ \citenamefont {Watanabe}}]{mariani:2010}%
  \BibitemOpen
  \bibfield  {author} {\bibinfo {author} {\bibfnamefont {G.}~\bibnamefont
  {Mariani}}, \bibinfo {author} {\bibfnamefont {L.}~\bibnamefont {Bruselli}},
  \bibinfo {author} {\bibfnamefont {T.}~\bibnamefont {Kuwert}}, \bibinfo
  {author} {\bibfnamefont {E.~E.}\ \bibnamefont {Kim}}, \bibinfo {author}
  {\bibfnamefont {A.}~\bibnamefont {Flotats}}, \bibinfo {author} {\bibfnamefont
  {O.}~\bibnamefont {Israel}}, \bibinfo {author} {\bibfnamefont
  {M.}~\bibnamefont {Dondi}}, \ and\ \bibinfo {author} {\bibfnamefont
  {N.}~\bibnamefont {Watanabe}},\ }\bibfield  {title} {\enquote {\bibinfo
  {title} {A review on the clinical uses of {SPECT/CT}},}\ }\href@noop {}
  {\bibfield  {journal} {\bibinfo  {journal} {Eur J Nucl Med Mol Imaging}\
  }\textbf {\bibinfo {volume} {37}},\ \bibinfo {pages} {1959--1985} (\bibinfo
  {year} {2010})}\BibitemShut {NoStop}%
\bibitem [{\citenamefont {Vavilov}(2015)}]{vavilov:2015}%
  \BibitemOpen
  \bibfield  {author} {\bibinfo {author} {\bibfnamefont {V.~P.}\ \bibnamefont
  {Vavilov}},\ }\bibfield  {title} {\enquote {\bibinfo {title} {Dynamic thermal
  tomography: Recent improvements and applications},}\ }\href@noop {}
  {\bibfield  {journal} {\bibinfo  {journal} {NDT \& E International}\ }\textbf
  {\bibinfo {volume} {71}},\ \bibinfo {pages} {23--32} (\bibinfo {year}
  {2015})}\BibitemShut {NoStop}%
\bibitem [{\citenamefont {Niu}\ \emph {et~al.}(2011)\citenamefont {Niu},
  \citenamefont {Yang}, \citenamefont {Jin}, \citenamefont {Wernick},\ and\
  \citenamefont {King}}]{niu:2011}%
  \BibitemOpen
  \bibfield  {author} {\bibinfo {author} {\bibfnamefont {X.}~\bibnamefont
  {Niu}}, \bibinfo {author} {\bibfnamefont {Y.}~\bibnamefont {Yang}}, \bibinfo
  {author} {\bibfnamefont {M.}~\bibnamefont {Jin}}, \bibinfo {author}
  {\bibfnamefont {M.~N.}\ \bibnamefont {Wernick}}, \ and\ \bibinfo {author}
  {\bibfnamefont {M.~A.}\ \bibnamefont {King}},\ }\bibfield  {title} {\enquote
  {\bibinfo {title} {Effects of motion, attenuation, and scatter corrections on
  gated cardiac {SPECT} reconstruction},}\ }\href@noop {} {\bibfield  {journal}
  {\bibinfo  {journal} {Med Phys}\ }\textbf {\bibinfo {volume} {38}},\ \bibinfo
  {pages} {6571--6584} (\bibinfo {year} {2011})}\BibitemShut {NoStop}%
\bibitem [{\citenamefont {Ritt}\ \emph {et~al.}(2011)\citenamefont {Ritt},
  \citenamefont {Vija}, \citenamefont {Hornegger},\ and\ \citenamefont
  {Kuwert}}]{ritt:2011}%
  \BibitemOpen
  \bibfield  {author} {\bibinfo {author} {\bibfnamefont {P.}~\bibnamefont
  {Ritt}}, \bibinfo {author} {\bibfnamefont {H.}~\bibnamefont {Vija}}, \bibinfo
  {author} {\bibfnamefont {J.}~\bibnamefont {Hornegger}}, \ and\ \bibinfo
  {author} {\bibfnamefont {T.}~\bibnamefont {Kuwert}},\ }\bibfield  {title}
  {\enquote {\bibinfo {title} {Absolute quantification in {SPECT}},}\
  }\href@noop {} {\bibfield  {journal} {\bibinfo  {journal} {Eur J Nucl Med Mol
  Imaging}\ }\textbf {\bibinfo {volume} {38}},\ \bibinfo {pages} {69--77}
  (\bibinfo {year} {2011})}\BibitemShut {NoStop}%
\bibitem [{\citenamefont {LaCroix}\ \emph {et~al.}(1994)\citenamefont
  {LaCroix}, \citenamefont {Tsui}, \citenamefont {Hasegawa},\ and\
  \citenamefont {Brown}}]{lacroix:1994}%
  \BibitemOpen
  \bibfield  {author} {\bibinfo {author} {\bibfnamefont {K.}~\bibnamefont
  {LaCroix}}, \bibinfo {author} {\bibfnamefont {B.}~\bibnamefont {Tsui}},
  \bibinfo {author} {\bibfnamefont {B.}~\bibnamefont {Hasegawa}}, \ and\
  \bibinfo {author} {\bibfnamefont {J.}~\bibnamefont {Brown}},\ }\bibfield
  {title} {\enquote {\bibinfo {title} {Investigation of the use of {X}-ray {CT}
  images for attenuation compensation in {SPECT}},}\ }\href@noop {} {\bibfield
  {journal} {\bibinfo  {journal} {IEEE Trans Nucl Sci}\ }\textbf {\bibinfo
  {volume} {41}},\ \bibinfo {pages} {2793--2799} (\bibinfo {year}
  {1994})}\BibitemShut {NoStop}%
\bibitem [{\citenamefont {Frey}\ and\ \citenamefont {Tsui}(1994)}]{Frey:1994}%
  \BibitemOpen
  \bibfield  {author} {\bibinfo {author} {\bibfnamefont {E.}~\bibnamefont
  {Frey}}\ and\ \bibinfo {author} {\bibfnamefont {B.}~\bibnamefont {Tsui}},\
  }\bibfield  {title} {\enquote {\bibinfo {title} {Modeling the scatter
  response function in inhomogeneous scattering media for {SPECT}},}\
  }\href@noop {} {\bibfield  {journal} {\bibinfo  {journal} {IEEE Trans Nucl
  Sci}\ }\textbf {\bibinfo {volume} {41}},\ \bibinfo {pages} {1585--1593}
  (\bibinfo {year} {1994})}\BibitemShut {NoStop}%
\bibitem [{\citenamefont {Koral}\ \emph {et~al.}(1988)\citenamefont {Koral},
  \citenamefont {Wang}, \citenamefont {Rogers}, \citenamefont {Clinthorne},\
  and\ \citenamefont {Wang}}]{Koral:1988}%
  \BibitemOpen
  \bibfield  {author} {\bibinfo {author} {\bibfnamefont {K.~F.}\ \bibnamefont
  {Koral}}, \bibinfo {author} {\bibfnamefont {X.}~\bibnamefont {Wang}},
  \bibinfo {author} {\bibfnamefont {W.~L.}\ \bibnamefont {Rogers}}, \bibinfo
  {author} {\bibfnamefont {N.~H.}\ \bibnamefont {Clinthorne}}, \ and\ \bibinfo
  {author} {\bibfnamefont {X.}~\bibnamefont {Wang}},\ }\bibfield  {title}
  {\enquote {\bibinfo {title} {{SPECT} compton-scattering correction by
  analysis of energy spectra},}\ }\href@noop {} {\bibfield  {journal} {\bibinfo
   {journal} {J Nucl Med}\ }\textbf {\bibinfo {volume} {29}},\ \bibinfo {pages}
  {195--202} (\bibinfo {year} {1988})}\BibitemShut {NoStop}%
\bibitem [{\citenamefont {Jaszczak}\ \emph {et~al.}(1984)\citenamefont
  {Jaszczak}, \citenamefont {Greer}, \citenamefont {Floyd}, \citenamefont
  {Harris},\ and\ \citenamefont {Coleman}}]{Jas:1984}%
  \BibitemOpen
  \bibfield  {author} {\bibinfo {author} {\bibfnamefont {R.~J.}\ \bibnamefont
  {Jaszczak}}, \bibinfo {author} {\bibfnamefont {K.~L.}\ \bibnamefont {Greer}},
  \bibinfo {author} {\bibfnamefont {J.~C.}\ \bibnamefont {Floyd}}, \bibinfo
  {author} {\bibfnamefont {C.~C.}\ \bibnamefont {Harris}}, \ and\ \bibinfo
  {author} {\bibfnamefont {R.~E.}\ \bibnamefont {Coleman}},\ }\bibfield
  {title} {\enquote {\bibinfo {title} {Improved {SPECT} quantification using
  compensation for scattered photons.}}\ }\href@noop {} {\bibfield  {journal}
  {\bibinfo  {journal} {J Nucl Med}\ }\textbf {\bibinfo {volume} {25}},\
  \bibinfo {pages} {893--900} (\bibinfo {year} {1984})}\BibitemShut {NoStop}%
\bibitem [{\citenamefont {Papanicolas}\ and\ \citenamefont
  {Stiliaris}(2012)}]{Papanicolas:2012}%
  \BibitemOpen
  \bibfield  {author} {\bibinfo {author} {\bibfnamefont {C.~N.}\ \bibnamefont
  {Papanicolas}}\ and\ \bibinfo {author} {\bibfnamefont {E.}~\bibnamefont
  {Stiliaris}},\ }\bibfield  {title} {\enquote {\bibinfo {title} {A novel
  method of data analysis for hadronic physics},}\ }\href@noop {} {\bibfield
  {journal} {\bibinfo  {journal} {arXiv: 1205.6505}\ } (\bibinfo {year}
  {2012})}\BibitemShut {NoStop}%
\bibitem [{\citenamefont {{L. Markou, E. Stiliaris and C. N.
  Papanicolas}}(2012)}]{Markou:2018}%
  \BibitemOpen
  \bibfield  {author} {\bibinfo {author} {\bibnamefont {{L. Markou, E.
  Stiliaris and C. N. Papanicolas}}},\ }\bibfield  {title} {\enquote {\bibinfo
  {title} {On hadron deformation: a model independent extraction of {EMR} from
  pion photoproduction data},}\ }\href@noop {} {\bibfield  {journal} {\bibinfo
  {journal} {arXiv: 1803.06283}\ } (\bibinfo {year} {2012})}\BibitemShut
  {NoStop}%
\bibitem [{\citenamefont {{C. Alexandrou, C. N. Papanicolas and M.
  Vanderhaeghen}}(2012)}]{Alexandrou:2012}%
  \BibitemOpen
  \bibfield  {author} {\bibinfo {author} {\bibnamefont {{C. Alexandrou, C. N.
  Papanicolas and M. Vanderhaeghen}}},\ }\bibfield  {title} {\enquote {\bibinfo
  {title} {{T}he {S}hape of {H}adrons},}\ }\href {\doibase
  10.1103/RevModPhys.84.1231} {\bibfield  {journal} {\bibinfo  {journal} {Rev
  Mod Phys}\ }\textbf {\bibinfo {volume} {84}},\ \bibinfo {pages} {1231--1251}
  (\bibinfo {year} {2012})}\BibitemShut {NoStop}%
\bibitem [{\citenamefont {Stiliaris}\ and\ \citenamefont
  {Papanicolas}(2007)}]{Stiliaris:2007}%
  \BibitemOpen
  \bibfield  {author} {\bibinfo {author} {\bibfnamefont {E.}~\bibnamefont
  {Stiliaris}}\ and\ \bibinfo {author} {\bibfnamefont {C.~N.}\ \bibnamefont
  {Papanicolas}},\ }\bibfield  {title} {\enquote {\bibinfo {title} {Multipole
  extraction: A novel, model independent method},}\ }in\ \href@noop {} {\emph
  {\bibinfo {booktitle} {AIP Conference Proceedings}}},\ Vol.\ \bibinfo
  {volume} {904}\ (\bibinfo {organization} {AIP},\ \bibinfo {year} {2007})\
  pp.\ \bibinfo {pages} {257--268}\BibitemShut {NoStop}%
\bibitem [{\citenamefont {Alexandrou}\ \emph {et~al.}(2015)\citenamefont
  {Alexandrou}, \citenamefont {Leontiou}, \citenamefont {Papanicolas},\ and\
  \citenamefont {Stiliaris}}]{Alexandrou:2015}%
  \BibitemOpen
  \bibfield  {author} {\bibinfo {author} {\bibfnamefont {C.}~\bibnamefont
  {Alexandrou}}, \bibinfo {author} {\bibfnamefont {T.}~\bibnamefont
  {Leontiou}}, \bibinfo {author} {\bibfnamefont {C.}~\bibnamefont
  {Papanicolas}}, \ and\ \bibinfo {author} {\bibfnamefont {E.}~\bibnamefont
  {Stiliaris}},\ }\bibfield  {title} {\enquote {\bibinfo {title} {Novel
  analysis method for excited states in lattice {QCD}: The nucleon case},}\
  }\href@noop {} {\bibfield  {journal} {\bibinfo  {journal} {Phys Rev D}\
  }\textbf {\bibinfo {volume} {91}},\ \bibinfo {pages} {014506} (\bibinfo
  {year} {2015})}\BibitemShut {NoStop}%
\bibitem [{\citenamefont {Epstein}(2007)}]{Epstein:2007}%
  \BibitemOpen
  \bibfield  {author} {\bibinfo {author} {\bibfnamefont {C.~L.}\ \bibnamefont
  {Epstein}},\ }\href@noop {} {\emph {\bibinfo {title} {Introduction to the
  mathematics of medical imaging}}}\ (\bibinfo  {publisher} {SIAM},\ \bibinfo
  {year} {2007})\BibitemShut {NoStop}%
\bibitem [{\citenamefont {Natterer}(2001)}]{Natterer:2001}%
  \BibitemOpen
  \bibfield  {author} {\bibinfo {author} {\bibfnamefont {F.}~\bibnamefont
  {Natterer}},\ }\bibfield  {title} {\enquote {\bibinfo {title} {Inversion of
  the attenuated {R}adon transform},}\ }\href@noop {} {\bibfield  {journal}
  {\bibinfo  {journal} {Inverse Probl}\ }\textbf {\bibinfo {volume} {17}},\
  \bibinfo {pages} {113} (\bibinfo {year} {2001})}\BibitemShut {NoStop}%
\bibitem [{\citenamefont {Novikov}(2002)}]{Novikov:2002}%
  \BibitemOpen
  \bibfield  {author} {\bibinfo {author} {\bibfnamefont {R.~G.}\ \bibnamefont
  {Novikov}},\ }\bibfield  {title} {\enquote {\bibinfo {title} {An inversion
  formula for the attenuated {X}-ray transformation},}\ }\href@noop {}
  {\bibfield  {journal} {\bibinfo  {journal} {Ark Mat}\ }\textbf {\bibinfo
  {volume} {40}},\ \bibinfo {pages} {145--167} (\bibinfo {year}
  {2002})}\BibitemShut {NoStop}%
\bibitem [{\citenamefont {Radon}(1917)}]{Radon:1917}%
  \BibitemOpen
  \bibfield  {author} {\bibinfo {author} {\bibfnamefont {J.}~\bibnamefont
  {Radon}},\ }\bibfield  {title} {\enquote {\bibinfo {title} {{\"U}ber die
  {B}estimmung von {F}unktionen durch ihre {I}ntegralwerte l{\"a}ngs gewisser
  {M}annigfaltigkeiten},}\ }\href@noop {} {\bibfield  {journal} {\bibinfo
  {journal} {Ber. Verh. Sachs. Akad. Wiss.}\ }\textbf {\bibinfo {volume}
  {69}},\ \bibinfo {pages} {262--277} (\bibinfo {year} {1917})}\BibitemShut
  {NoStop}%
\bibitem [{\citenamefont {Bruyant}(2002)}]{bruyant:2002}%
  \BibitemOpen
  \bibfield  {author} {\bibinfo {author} {\bibfnamefont {P.~P.}\ \bibnamefont
  {Bruyant}},\ }\bibfield  {title} {\enquote {\bibinfo {title} {Analytic and
  iterative reconstruction algorithms in {SPECT}},}\ }\href@noop {} {\bibfield
  {journal} {\bibinfo  {journal} {J Nucl Med}\ }\textbf {\bibinfo {volume}
  {43}},\ \bibinfo {pages} {1343--1358} (\bibinfo {year} {2002})}\BibitemShut
  {NoStop}%
\bibitem [{\citenamefont {Qi}\ and\ \citenamefont {Leahy}(2006)}]{qi:2006}%
  \BibitemOpen
  \bibfield  {author} {\bibinfo {author} {\bibfnamefont {J.}~\bibnamefont
  {Qi}}\ and\ \bibinfo {author} {\bibfnamefont {R.~M.}\ \bibnamefont {Leahy}},\
  }\bibfield  {title} {\enquote {\bibinfo {title} {Iterative reconstruction
  techniques in emission computed tomography},}\ }\href@noop {} {\bibfield
  {journal} {\bibinfo  {journal} {Phys Med Biol}\ }\textbf {\bibinfo {volume}
  {51}},\ \bibinfo {pages} {R541} (\bibinfo {year} {2006})}\BibitemShut
  {NoStop}%
\bibitem [{\citenamefont {Gordon}\ \emph {et~al.}(1970)\citenamefont {Gordon}
  \emph {et~al.}}]{Gordon:1970}%
  \BibitemOpen
  \bibfield  {author} {\bibinfo {author} {\bibfnamefont {R.}~\bibnamefont
  {Gordon}} \emph {et~al.},\ }\bibfield  {title} {\enquote {\bibinfo {title}
  {Algebraic reconstruction techniques ({ART}) for three-dimensional electron
  microscopy and {X}-ray photography},}\ }\href@noop {} {\bibfield  {journal}
  {\bibinfo  {journal} {J Theor Biol}\ }\textbf {\bibinfo {volume} {29}},\
  \bibinfo {pages} {471 -- 481} (\bibinfo {year} {1970})}\BibitemShut {NoStop}%
\bibitem [{\citenamefont {Gilbert}(1972)}]{Gilbert:1972}%
  \BibitemOpen
  \bibfield  {author} {\bibinfo {author} {\bibfnamefont {P.}~\bibnamefont
  {Gilbert}},\ }\bibfield  {title} {\enquote {\bibinfo {title} {Iterative
  methods for the three-dimensional reconstruction of an object from
  projections},}\ }\href@noop {} {\bibfield  {journal} {\bibinfo  {journal} {J
  Theor Biol}\ }\textbf {\bibinfo {volume} {36}},\ \bibinfo {pages} {105 --
  117} (\bibinfo {year} {1972})}\BibitemShut {NoStop}%
\bibitem [{\citenamefont {Shepp}\ and\ \citenamefont
  {Vardi}(1982)}]{Shepp:1982}%
  \BibitemOpen
  \bibfield  {author} {\bibinfo {author} {\bibfnamefont {L.~A.}\ \bibnamefont
  {Shepp}}\ and\ \bibinfo {author} {\bibfnamefont {Y.}~\bibnamefont {Vardi}},\
  }\bibfield  {title} {\enquote {\bibinfo {title} {Maximum {L}ikelihood
  {R}econstruction for {E}mission {T}omography},}\ }\href@noop {} {\bibfield
  {journal} {\bibinfo  {journal} {IEEE Trans Med Imag}\ }\textbf {\bibinfo
  {volume} {1}},\ \bibinfo {pages} {113--122} (\bibinfo {year}
  {1982})}\BibitemShut {NoStop}%
\bibitem [{\citenamefont {Lange}\ and\ \citenamefont
  {Carson}(1984)}]{Lange:1984}%
  \BibitemOpen
  \bibfield  {author} {\bibinfo {author} {\bibfnamefont {K.}~\bibnamefont
  {Lange}}\ and\ \bibinfo {author} {\bibfnamefont {R.}~\bibnamefont {Carson}},\
  }\bibfield  {title} {\enquote {\bibinfo {title} {{EM} reconstruction
  algorithms for emission and transmission tomography},}\ }\href@noop {}
  {\bibfield  {journal} {\bibinfo  {journal} {J Comput Assist Tomogr}\ }\textbf
  {\bibinfo {volume} {8}},\ \bibinfo {pages} {306--16} (\bibinfo {year}
  {1984})}\BibitemShut {NoStop}%
\bibitem [{\citenamefont {Hudson}\ and\ \citenamefont
  {Larkin}(1994)}]{hudson:1994}%
  \BibitemOpen
  \bibfield  {author} {\bibinfo {author} {\bibfnamefont {H.~M.}\ \bibnamefont
  {Hudson}}\ and\ \bibinfo {author} {\bibfnamefont {R.~S.}\ \bibnamefont
  {Larkin}},\ }\bibfield  {title} {\enquote {\bibinfo {title} {Accelerated
  image reconstruction using ordered subsets of projection data},}\ }\href@noop
  {} {\bibfield  {journal} {\bibinfo  {journal} {IEEE Trans Med Imag}\ }\textbf
  {\bibinfo {volume} {13}},\ \bibinfo {pages} {601--609} (\bibinfo {year}
  {1994})}\BibitemShut {NoStop}%
\bibitem [{\citenamefont {Wieczorek}(2010)}]{Wiez:2010}%
  \BibitemOpen
  \bibfield  {author} {\bibinfo {author} {\bibfnamefont {H.}~\bibnamefont
  {Wieczorek}},\ }\bibfield  {title} {\enquote {\bibinfo {title} {The image
  quality of {FBP} and {MLEM} reconstruction},}\ }\href@noop {} {\bibfield
  {journal} {\bibinfo  {journal} {Phys Med Biol}\ }\textbf {\bibinfo {volume}
  {55}},\ \bibinfo {pages} {3161} (\bibinfo {year} {2010})}\BibitemShut
  {NoStop}%
\bibitem [{\citenamefont {Ma}\ \emph {et~al.}(2013)\citenamefont {Ma},
  \citenamefont {Wolf}, \citenamefont {Clough},\ and\ \citenamefont
  {Schmidt}}]{ma:2013}%
  \BibitemOpen
  \bibfield  {author} {\bibinfo {author} {\bibfnamefont {D.}~\bibnamefont
  {Ma}}, \bibinfo {author} {\bibfnamefont {P.}~\bibnamefont {Wolf}}, \bibinfo
  {author} {\bibfnamefont {A.~V.}\ \bibnamefont {Clough}}, \ and\ \bibinfo
  {author} {\bibfnamefont {T.~G.}\ \bibnamefont {Schmidt}},\ }\bibfield
  {title} {\enquote {\bibinfo {title} {The performance of {MLEM} for dynamic
  imaging from simulated few-view, multi-pinhole {SPECT}},}\ }\href@noop {}
  {\bibfield  {journal} {\bibinfo  {journal} {IEEE Trans Nucl Sci}\ }\textbf
  {\bibinfo {volume} {60}},\ \bibinfo {pages} {115--123} (\bibinfo {year}
  {2013})}\BibitemShut {NoStop}%
\bibitem [{\citenamefont {Wolf}\ \emph {et~al.}(2013)\citenamefont {Wolf},
  \citenamefont {J{\o}rgensen}, \citenamefont {Schmidt},\ and\ \citenamefont
  {Sidky}}]{wolf:2013}%
  \BibitemOpen
  \bibfield  {author} {\bibinfo {author} {\bibfnamefont {P.~A.}\ \bibnamefont
  {Wolf}}, \bibinfo {author} {\bibfnamefont {J.~S.}\ \bibnamefont
  {J{\o}rgensen}}, \bibinfo {author} {\bibfnamefont {T.~G.}\ \bibnamefont
  {Schmidt}}, \ and\ \bibinfo {author} {\bibfnamefont {E.~Y.}\ \bibnamefont
  {Sidky}},\ }\bibfield  {title} {\enquote {\bibinfo {title} {Few-view single
  photon emission computed tomography ({SPECT}) reconstruction based on a
  blurred piecewise constant object model},}\ }\href@noop {} {\bibfield
  {journal} {\bibinfo  {journal} {Phys Med Biol}\ }\textbf {\bibinfo {volume}
  {58}},\ \bibinfo {pages} {5629} (\bibinfo {year} {2013})}\BibitemShut
  {NoStop}%
\bibitem [{\citenamefont {Angeli}\ and\ \citenamefont
  {Stiliaris}(2009)}]{Angeli:2009}%
  \BibitemOpen
  \bibfield  {author} {\bibinfo {author} {\bibfnamefont {S.}~\bibnamefont
  {Angeli}}\ and\ \bibinfo {author} {\bibfnamefont {E.}~\bibnamefont
  {Stiliaris}},\ }\bibfield  {title} {\enquote {\bibinfo {title} {An
  accelerated algebraic reconstruction technique based on the
  {N}ewton-{R}aphson scheme},}\ }\href@noop {} {\bibfield  {journal} {\bibinfo
  {journal} {2009 IEEE NSS/MIC}\ ,\ \bibinfo {pages} {3382--3387}} (\bibinfo
  {year} {2009})}\BibitemShut {NoStop}%
\bibitem [{\citenamefont {G{\"u}rsoy}\ \emph {et~al.}(2014)\citenamefont
  {G{\"u}rsoy}, \citenamefont {De~Carlo}, \citenamefont {Xiao},\ and\
  \citenamefont {Jacobsen}}]{gursoy:2014}%
  \BibitemOpen
  \bibfield  {author} {\bibinfo {author} {\bibfnamefont {D.}~\bibnamefont
  {G{\"u}rsoy}}, \bibinfo {author} {\bibfnamefont {F.}~\bibnamefont
  {De~Carlo}}, \bibinfo {author} {\bibfnamefont {X.}~\bibnamefont {Xiao}}, \
  and\ \bibinfo {author} {\bibfnamefont {C.}~\bibnamefont {Jacobsen}},\
  }\bibfield  {title} {\enquote {\bibinfo {title} {{TomoPy}: a framework for
  the analysis of synchrotron tomographic data},}\ }\href@noop {} {\bibfield
  {journal} {\bibinfo  {journal} {J. Synchrotron Radiat}\ }\textbf {\bibinfo
  {volume} {21}},\ \bibinfo {pages} {1188--1193} (\bibinfo {year}
  {2014})}\BibitemShut {NoStop}%
\bibitem [{\citenamefont {Lencrerot}\ \emph {et~al.}(2009)\citenamefont
  {Lencrerot}, \citenamefont {Litman}, \citenamefont {Tortel},\ and\
  \citenamefont {Geffrin}}]{len:2009}%
  \BibitemOpen
  \bibfield  {author} {\bibinfo {author} {\bibfnamefont {R.}~\bibnamefont
  {Lencrerot}}, \bibinfo {author} {\bibfnamefont {A.}~\bibnamefont {Litman}},
  \bibinfo {author} {\bibfnamefont {H.}~\bibnamefont {Tortel}}, \ and\ \bibinfo
  {author} {\bibfnamefont {J.-M.}\ \bibnamefont {Geffrin}},\ }\bibfield
  {title} {\enquote {\bibinfo {title} {Imposing zernike representation for
  imaging two-dimensional targets},}\ }\href@noop {} {\bibfield  {journal}
  {\bibinfo  {journal} {Inverse Probl}\ }\textbf {\bibinfo {volume} {25}},\
  \bibinfo {pages} {035012} (\bibinfo {year} {2009})}\BibitemShut {NoStop}%
\bibitem [{\citenamefont {Born}\ and\ \citenamefont {Wolf}(1959)}]{born:1959}%
  \BibitemOpen
  \bibfield  {author} {\bibinfo {author} {\bibfnamefont {M.}~\bibnamefont
  {Born}}\ and\ \bibinfo {author} {\bibfnamefont {E.}~\bibnamefont {Wolf}},\
  }\href@noop {} {\emph {\bibinfo {title} {Principles of optics}}}\ (\bibinfo
  {publisher} {New York: Pergamon},\ \bibinfo {year} {1959})\BibitemShut
  {NoStop}%
\bibitem [{\citenamefont {Schwarz}\ \emph {et~al.}(1978)\citenamefont {Schwarz}
  \emph {et~al.}}]{schwarz:1978}%
  \BibitemOpen
  \bibfield  {author} {\bibinfo {author} {\bibfnamefont {G.}~\bibnamefont
  {Schwarz}} \emph {et~al.},\ }\bibfield  {title} {\enquote {\bibinfo {title}
  {Estimating the dimension of a model},}\ }\href@noop {} {\bibfield  {journal}
  {\bibinfo  {journal} {Annals Stat}\ }\textbf {\bibinfo {volume} {6}},\
  \bibinfo {pages} {461--464} (\bibinfo {year} {1978})}\BibitemShut {NoStop}%
\bibitem [{\citenamefont {Neath}\ and\ \citenamefont
  {Cavanaugh}(2012)}]{neath:2012}%
  \BibitemOpen
  \bibfield  {author} {\bibinfo {author} {\bibfnamefont {A.~A.}\ \bibnamefont
  {Neath}}\ and\ \bibinfo {author} {\bibfnamefont {J.~E.}\ \bibnamefont
  {Cavanaugh}},\ }\bibfield  {title} {\enquote {\bibinfo {title} {The
  {B}ayesian {I}nformation {C}riterion: background, derivation, and
  applications},}\ }\href@noop {} {\bibfield  {journal} {\bibinfo  {journal}
  {Wiley Interdiscip Rev Comput Stat.}\ }\textbf {\bibinfo {volume} {4}},\
  \bibinfo {pages} {199--203} (\bibinfo {year} {2012})}\BibitemShut {NoStop}%
\bibitem [{\citenamefont {Hill}\ \emph {et~al.}(2001)\citenamefont {Hill} \emph
  {et~al.}}]{Hill:2001}%
  \BibitemOpen
  \bibfield  {author} {\bibinfo {author} {\bibfnamefont {D.~L.~G.}\
  \bibnamefont {Hill}} \emph {et~al.},\ }\bibfield  {title} {\enquote {\bibinfo
  {title} {Medical image registration},}\ }\href@noop {} {\bibfield  {journal}
  {\bibinfo  {journal} {Phys Med Biol}\ }\textbf {\bibinfo {volume} {46}},\
  \bibinfo {pages} {R1} (\bibinfo {year} {2001})}\BibitemShut {NoStop}%
\bibitem [{\citenamefont {Koutsantonis}, \citenamefont {Lemesios},\ and\
  \citenamefont {Papanicolas}(2017)}]{koutsantonis:2017}%
  \BibitemOpen
  \bibfield  {author} {\bibinfo {author} {\bibfnamefont {L.}~\bibnamefont
  {Koutsantonis}}, \bibinfo {author} {\bibfnamefont {C.}~\bibnamefont
  {Lemesios}}, \ and\ \bibinfo {author} {\bibfnamefont {C.~N.}\ \bibnamefont
  {Papanicolas}},\ }\bibfield  {title} {\enquote {\bibinfo {title} {A
  reconstruction method based on a data analysis scheme for {SPECT} imaging in
  {P}arkinson's disease},}\ }\href@noop {} {\bibfield  {journal} {\bibinfo
  {journal} {2017 IEEE NSS/MIC/RTSD}\ } (\bibinfo {year} {2017})}\BibitemShut
  {NoStop}%
\bibitem [{\citenamefont {Koutsantonis}(2017)}]{koutsantonis:2017mouse}%
  \BibitemOpen
  \bibfield  {author} {\bibinfo {author} {\bibfnamefont {L.}~\bibnamefont
  {Koutsantonis}},\ }\href@noop {} {\enquote {\bibinfo {title} {{AMIAS} image
  reconstruction - {A}pplication in small animal imaging with {SPECT}},}\
  }\bibinfo {type} {Tech. Rep.}\ (\bibinfo  {institution} {The Cyprus
  Institute},\ \bibinfo {year} {2017})\BibitemShut {NoStop}%
\bibitem [{\citenamefont {Spanoudaki}\ \emph {et~al.}(2004)\citenamefont
  {Spanoudaki}, \citenamefont {Giokaris}, \citenamefont {Karabarbounis} \emph
  {et~al.}}]{Spanoudaki:2004}%
  \BibitemOpen
  \bibfield  {author} {\bibinfo {author} {\bibfnamefont {V.}~\bibnamefont
  {Spanoudaki}}, \bibinfo {author} {\bibfnamefont {N.}~\bibnamefont
  {Giokaris}}, \bibinfo {author} {\bibfnamefont {A.}~\bibnamefont
  {Karabarbounis}},  \emph {et~al.},\ }\bibfield  {title} {\enquote {\bibinfo
  {title} {Design and development of a position-sensitive $\gamma$-camera for
  {SPECT} imaging based on {PCI} electronics},}\ }\href@noop {} {\bibfield
  {journal} {\bibinfo  {journal} {Nucl Instr Meth Phys Res A}\ }\textbf
  {\bibinfo {volume} {527}},\ \bibinfo {pages} {151--156} (\bibinfo {year}
  {2004})}\BibitemShut {NoStop}%
\bibitem [{\citenamefont {Koutsantonis}\ \emph {et~al.}(2016)\citenamefont
  {Koutsantonis}, \citenamefont {Papanicolas}, \citenamefont {Rapsomanikis},\
  and\ \citenamefont {Stiliaris}}]{koutsantonis:2016}%
  \BibitemOpen
  \bibfield  {author} {\bibinfo {author} {\bibfnamefont {L.}~\bibnamefont
  {Koutsantonis}}, \bibinfo {author} {\bibfnamefont {C.~N.}\ \bibnamefont
  {Papanicolas}}, \bibinfo {author} {\bibfnamefont {A.-N.}\ \bibnamefont
  {Rapsomanikis}}, \ and\ \bibinfo {author} {\bibfnamefont {E.}~\bibnamefont
  {Stiliaris}},\ }\bibfield  {title} {\enquote {\bibinfo {title} {{AMIAS}: A
  novel statistical method for tomographic image reconstruction - {A}pplication
  in thermal emission tomography},}\ }\href@noop {} {\bibfield  {journal}
  {\bibinfo  {journal} {2016 IEEE NSS/MIC/RTSD}\ } (\bibinfo {year}
  {2016})}\BibitemShut {NoStop}%
\end{thebibliography}

%

\end{document}